\documentclass[twocolumn,amsmath,amssymb,superscriptaddress]{revtex4-1}
\usepackage{graphicx}% Include figure files
\usepackage{color}
\usepackage[caption=false]{subfig}

\def\figdir{.}
\begin{document}

\title{Combining density based dynamical correlation with a reduced density matrix strong correlation description} 

\author{Robert van Meer} 
\email[Corresponding Author. Email: ]{rvanmeer@gmail.com} 
\affiliation{Department of Physics, National Taiwan University, Taipei 10617, Taiwan} 
\author{Oleg Gritsenko} 
\affiliation{Section Theoretical Chemistry, VU University, NL-1081 HV Amsterdam, The Netherlands}
\affiliation{Institute of Physics, Lodz University of Technology, PL-90-924 Lodz, Poland}
\author{Jeng-Da Chai} 
\email[Corresponding Author. Email: ]{jdchai@phys.ntu.edu.tw} 
\affiliation{Department of Physics, National Taiwan University, Taipei 10617, Taiwan} 
\affiliation{Center for Theoretical Physics and Center for Quantum Science and Engineering, National Taiwan University, Taipei 10617, Taiwan} 

\date{\today} 

\begin{abstract} 
\noindent 
A combined density and first-order reduced-density-matrix (1RDM) functional method is proposed for the calculation of potential energy curves (PECs) of molecular multibond dissociation. Its 1RDM functional part, a pair density functional, efficiently approximates the {\it ab initio} pair density of the complete active space self-consistent-field (CASSCF) method. The corresponding approximate on top pair density $\Pi$ is employed to correct for double counting a correlation functional of density functional theory (DFT). The proposed ELS-DM$\Pi$DFT method with the extended L\"{o}wdin-Shull (ELS) 1RDM functional with dispersion and multibond (DM) corrections augmented with the $\Pi$DFT functional closely reproduces PECs of multibond dissociation in the paradigmatic N$_2$, H$_2$O, and H$_2$CO molecules calculated with the recently proposed CAS$\Pi$DFT (CASSCF augmented with a $\Pi$ based scaled DFT correlation correction) method. Furthermore, with the additional M-correction, ELS-DM$\Pi$DFT+M reproduces well the benchmark PEC of the N$_2$ molecule by Lie and Clementi. 
\end{abstract} 

\maketitle

\section{Introduction} 
\noindent 
The adequate description of bond breaking processes often requires the correct handling of both dynamical and strong non-dynamical correlation at all bond distances. Conventional density functional theory (DFT) 
(i.e., Kohn-Sham DFT) \cite{HohenbergKohn1964,KohnSham1965} employing approximate density functionals is fully capable of handling the (mainly) dynamical correlation for equilibrium geometry structures, but 
fails to deliver an adequate description of the strong correlation that is required when one dissociates bonds \cite{Cohen2008b,CohenMoriYang2012ChRev}. Recently, thermally-assisted-occupation density 
functional theory (TAO-DFT) \cite{Chai2012,Chai2014,Chai2017}, an efficient method to describe both dynamical and strong non-dynamical correlation \cite{TAOa1,TAOa2}, has been developed. However, the 
choice of the fictitious temperature in TAO-DFT remains difficult, especially for molecular multibond dissociation \cite{LinHuiChungChai2017}. Alternatively, an effective way of describing the strong correlation using 
a functional like description is to use density matrix functional theory (DMFT). However, recent geminal based approximate DMFT functionals fail to describe 50-80 \% of the dynamical correlation and there does not 
seem to be a way to tackle this problem in a fully self-consistent fashion. In this paper we look at the previously developed CAS$\Pi$DFT method for guidance, and try to combine both functional approaches in order 
to obtain a functional based method that combines the best of both worlds and can generate rather accurate potential energy surfaces. 

In the CAS$\Pi$DFT method the electronic energy of a state is expressed in terms of the CASSCF energy $E_e^{CASSCF}$ and the $\Pi$DFT component $E^{\Pi DFT}$. The latter accounts for the dynamical correlation part which is not described by the CASSCF wavefunction
\begin{equation}
E_e^{CAS\Pi DFT} = E_e^{CASSCF} + E^{\Pi DFT}[X^{CASSCF}, \rho^{CASSCF}]
\end{equation}
The $\Pi$DFT component itself is generated by using a scaled the correlation energy density functional by Lee, Yang, and Parr (LYP) \cite{LeeYangParr1988} 
\begin{equation}
E^{\Pi DFT}[X,\rho] = \int P[x]\epsilon_c^{LYP}[\rho(\boldsymbol{r})]d\boldsymbol{r}
\end{equation}
whose scaling factor $P[X]$ depends on the on top density (pair density $\Pi(\boldsymbol{r}_1,\boldsymbol{r}_2)$ evaluated at $\boldsymbol{r}_1 = \boldsymbol{r}_2$ \cite{Gori-Giorgi2006,CarlsonTruhlarGagliardi2017,GritsenkovanMeerPernal2018,FerteGinerToulouse2019,AlcobaTorreLain2020}) and density $\rho$
\begin{align}
X(\boldsymbol{r}) = \frac{2\Pi(\boldsymbol{r},\boldsymbol{r})}{\rho(\boldsymbol{r})^2}
\end{align}
The currently used parametrization differentiates between two regions based on physical characteristics
\begin{equation}
P[X] =
\begin{cases}
P^{SDC}(X) \leq 1,  X \leq 1 \\
P^{EDC}(X) > 1, X > 1
\end{cases}
\end{equation} 
In case ($X\leq 1$) one is dealing with suppressed dynamical correlation (SDC), which is a situation that commonly occurs when bonds are being broken. The other scenario ($X > 1$) mainly occurs in energetically important spatial regions when one is describing distributed ionic type states, such as the first $^1\Sigma_u^+$ state of the H$_2$ molecule. In this case an enhanced dynamical correlation (EDC) description is warranted.

The CAS$\Pi$DFT method with its suppression and enhancement of dynamical correlation has been applied successfully to various ground and excited state systems \cite{GritsenkovanMeerPernal2018,GritsenkovanMeerPernal2019,GritsenkoPernal2019}. The latest variants, CAS$\Pi$DFT+M and CAS(M)$\Pi$DFT which also include an additional medium distance correlation correction, have been able to fairly accurately reproduce complete basis set (CBS) limit potential energy curves (PECs) for multibonded molecules \cite{PernalGritsenkovanMeer2019,HapkaPernalGritsenko2020}.

Up until this point the $\Pi$DFT scheme has always been used in conjunction with a relatively complicated wavefunction based CAS type (SCF or non-SCF) non-dynamical correlation carrier. In case of ground states one can also consider to use a less complicated functional based approach for obtaining the CASSCF energy, pair density and the comitant on top density and density quantities. The most suitable candidate is DMFT. 

In this paper we combine the DMFT approach with the $\Pi$DFT dynamical correlation correction. Section II describes the methodological details of the density matrix functional that is used for all calculations, extended L\"{o}wdin-Shull (ELS) with dispersion and multibond corrections (DM), and its utilization of the $\Pi$DFT correction scheme. In section III the full computational details of this endeavor are given. Section IV describes the application of the combined ELS-DM$\Pi$DFT scheme to several prototypical H$_2$O, N$_2$ and H$_2$CO molecules and compares the results with the CASSCF + $\Pi DFT$ and CBS benchmark data. Conclusions are drawn in the final section. 

\section{Density Matrix Functional Theory and the $\Pi$DFT correction} 
In DMFT the electronic ground state energy can be written as a functional of the one body reduced density matrix $\gamma(\boldsymbol{x},\boldsymbol{x}')$ \cite{Gilbert1975}, 
\begin{align}
E_e^{DMFT} [\gamma(\boldsymbol{x},\boldsymbol{x}')] = -\frac{1}{2}\int\nabla^2_{\boldsymbol{r}'}\gamma(\boldsymbol{x},\boldsymbol{x}')|_{\boldsymbol{x}'=\boldsymbol{x}}d\boldsymbol{x} \nonumber \\ + \int v_{\text{ext}}\gamma(\boldsymbol{x},\boldsymbol{x}')|_{\boldsymbol{x}'=\boldsymbol{x}}d\boldsymbol{x} + W^{\text{DMFT}}[\gamma(\boldsymbol{x},\boldsymbol{x}')]
\end{align}
Here $\boldsymbol{x}$ stands for the combination of the spatial $\boldsymbol{r}$ and spin s electron coordinates, and $W_{\text{DMFT}}[\gamma(\boldsymbol{x},\boldsymbol{x}')]$ is the two-electron interaction functional, whose exact form is only known for systems consisting of two electrons, requiring one to use approximate functionals for other systems. Several approximate functionals have been developed over the course of many years \cite{CsanyiArias2000,Muller1984,BuijseBaerends2002,CohenBaerends2002,GritsenkoPernalBaerends2005,RohrAC32008,LathiotakisSharma2009,ScuseriaTsuchimochi2009,Piris2014,Piris2017,vanMeerGritsenkoBaerends2018,vanMeerGritsenko2019}. 
All of these functionals can essentially be written as an integral over the approximate pair density in the natural orbital (NO) basis whose elements/prefactors are determined by one-body density matrix quantities.
\begin{align}
W^{\text{DMFT}}[\gamma(\boldsymbol{x},\boldsymbol{x}')] = \int d\boldsymbol{r}_1 d\boldsymbol{r}_2 \frac{\Pi^{\text{DMFT}}[\gamma(\boldsymbol{x},\boldsymbol{x}')](\boldsymbol{r}_1,\boldsymbol{r}_2)}{|\boldsymbol{r}_1-\boldsymbol{r}_2|}
\end{align}
The best candidate functional for our case is the ELS-DM functional, since it has been shown that this functional is fully capable of reproducing small CASSCF wavefunction results for small molecules \cite{vanMeerGritsenkoBaerends2018}. This functional is essentially an anti-symmetrized product of strongly orthogonal geminals (APSG) functional with additional dispersive dynamical correlation (D) and multibond dissociation (M) corrections.
\begin{align}
\Pi^{\text{ELS-DM}}(\boldsymbol{r}_1,\boldsymbol{r}_2) = \Pi^{\text{APSG}}(\boldsymbol{r}_1,\boldsymbol{r}_2) + \Pi^{\text{D}}(\boldsymbol{r}_1,\boldsymbol{r}_2) + \Pi^{\text{M}}(\boldsymbol{r}_1,\boldsymbol{r}_2) 
\end{align}
The APSG functional \cite{Rassolov2002,Pernal2013} divides the system into multiple two electron subsystems with their own set of NOs and uses the exact two electron L\"{o}wdin-Shull (LS) functional for the interaction of the orbitals (and electrons) within the set \cite{LowdinShull1956}, and a Hartree-Fock (HF) type interaction (no correlation) between orbitals belonging to different sets
\begin{align}
\Pi^{\text{APSG}}(\boldsymbol{r}_1,\boldsymbol{r}_2) = \sum_{i\in P}\sum_{j\in Q\neq P} n_in_j(4j_{ij}(\boldsymbol{r}_1,\boldsymbol{r}_2)-2k_{ij}(\boldsymbol{r}_1,\boldsymbol{r}_2)) \nonumber \\+ \sum_{i\in P}\sum_{j\in Q = P} f_if_j\sqrt{n_in_j}l_{ij}(\boldsymbol{r}_1,\boldsymbol{r}_2) 
\end{align}
Here P and Q denote sets, $f_i$ are the phase factors that have a value of 1 for the first member of a set and generally -1 for all other members of the set, $\phi_i(\boldsymbol{r})$ are the NOs and $n_i$ are the natural occupation numbers (NONs) whose value ranges from 0 to 1. The Coulomb $j_{ij}$, exchange $k_{ij}$, and star swapped exchange $l_{ij}$ orbital products lead to their respective integrals when integrated, and are given by 
\begin{align}
\label{eq:JKLij}
j_{ij}(\boldsymbol{r}_1,\boldsymbol{r}_2) =  \phi_i^*(\boldsymbol{r}_1)\phi_j^*(\boldsymbol{r}_2)\phi_i(\boldsymbol{r}_1)\phi_j(\boldsymbol{r}_2) \\
k_{ij}(\boldsymbol{r}_1,\boldsymbol{r}_2) =  \phi_i^*(\boldsymbol{r}_1)\phi_j^*(\boldsymbol{r}_2)\phi_j(\boldsymbol{r}_1)\phi_i(\boldsymbol{r}_2) \\
l_{ij}(\boldsymbol{r}_1,\boldsymbol{r}_2) =  \phi_i^*(\boldsymbol{r}_1)\phi_j(\boldsymbol{r}_2)\phi_j(\boldsymbol{r}_1)\phi_i^*(\boldsymbol{r}_2)
\end{align}
Note that the difference between the $k_{ij}$ and $l_{ij}$ orbital products only plays a role for the time dependent treatment \cite{Pernal2007a,PernalCioslowski2007,GiesbertzGritsenkoBaerends2012,vanMeerGritsenkoBaerends2014}, and is not important for the rest of this paper. The occupation numbers of the APSG functional follow a strict sum rule
\begin{align}
\sum_{i\in P} n_i = 1
\end{align}
ensuring that each geminal/set contains exactly two electrons (total occupation of 1 in our notation). While one can in principle assign any number of orbitals to a given geminal, one often resorts to only assigning two orbitals to every geminal (perfect pairing) due to the ambiguity of the assignment of more orbitals and the often relatively little energetic gain when doing so.

The $\Pi^M$ correction to the ELS-D functional represents the contribution to $\Pi$ due to non-dynamical correlation of electrons of different dissociating bonds 
\begin{align}
\label{eq:multibond}
&\Pi^{\text{M}}(\boldsymbol{r}_1,\boldsymbol{r}_2) = -\sum_{i\in P}\sum_{j\in Q\neq P} k_{ij} \nonumber \\ &P_m(n_i(1-n_i))P_m(n_j(1-n_j))\sqrt{n_i(1-n_i)n_j(1-n_j)} 
\end{align}
Here
\begin{equation}
P_m(z) = (1+\frac{16}{\gamma})\frac{\gamma z^2}{1+\gamma z^2} 
\end{equation} 
and $\gamma$ is a parameter. The $\Pi^{\text{D}}$ correction describes the dynamical dispersion type correlation between electrons on different geminals, and is responsible for up to 50 \% of the CASSCF equilibrium geometry correlation in case one uses an active space of 1 orbital for each valence electron 
\begin{align}
&\Pi^{\text{D}}(\boldsymbol{r}_1,\boldsymbol{r}_2) = \frac{1}{2} \sum_{i\neq a \in P} \sum_{j\neq b \in Q \neq P} F_D(n_i,n_a,n_jk,n_b) d_{ia,jb}
\end{align}
here the dispersive type orbital product is given by 
\begin{align}
d_{ia,jb}(\boldsymbol{r}_1,\boldsymbol{r}_2) = \phi_i(\boldsymbol{r}_1)\phi_j(\boldsymbol{r}_2)\phi_a^*(\boldsymbol{r}_1)\phi_b^*(\boldsymbol{r}_2)
\end{align}
and the $F_D$ prefactor is given by 
\begin{align}
F_D(n_i,n_a,n_j,n_b) = 8 f_{iajb} P_d(n_in_a)P_d(n_jn_b)\sqrt{n_in_an_jn_b} 
\end{align}
where $f_{iajb}$ are phase factors that ensure that the energetic contribution of each index combination is negative (attractive). In (17) $P_d(z)$ are the following functions of the NON products
\begin{align}
P_d(z) = \alpha \left ( 1 - \frac{\beta z^2}{1 + \beta z^2} \right ) 
\end{align}
with $\alpha$ and $\beta$ being the parameters (See below).

When both corrections are applied one can reproduce the energies of small CASSCF expansions. One should keep in mind that these CASSCF expansions still only cover 50 \% of the dynamical correlation, the other half of this correlation can only be captured by somehow incorporating the correlation space of the remainder of the complete set of "virtual" orbitals. As already mentioned before, in case of the APSG functional this additional space can be quite hard (and pointless) to incorporate. One can often only get relatively little energetic gain, while the efficiency of the SCF process is slowed down significantly by the constant moving of orbitals between sets. The situation improves slightly when additional intergeminal correlation is introduced. However, the higher lying virtuals can still not be assigned to a specific set. 

The most practical way to solve this issue is to use a method that does not require the set assignment of this "sea of virtuals". There are essentially two main classes of these general dynamical correlation schemes that can be used: perturbative approaches and scaled DFT correlation energy functionals. Several perturbative approaches have been successfully applied to geminal based functionals \cite{JeszenskiNagySurjan2014,ChatterjeePernal2016,Piris2017}, the main downside being the relatively large dependence on the size of the basis required for the proper account of dynamical correlation. Note, that DFT based approaches have a much smaller dependence on the basis set size, since they do not use unoccupied virtual orbitals. In our case we use such a DFT based approach and obtain the missing dynamical correlation by inserting the approximate ELS-DM on top pair density of Eq.(7) and density into the $\Pi$DFT expression (2) 
\begin{align}
E_e^{ELSDM\Pi DFT} =& E_e^{ELS-DM} + E^{\Pi DFT}[X^{ELS-DM}, \rho^{ELS-DM}] 
\end{align}
with
\begin{align}
X^{ELS-DM}(\boldsymbol{r}) =& \frac{2\Pi^{\text{ELS-DM}}(\boldsymbol{r},\boldsymbol{r})}{\rho^{\text{ELS-DM}}(\boldsymbol{r})^2}
\end{align}
resulting in a method that is completely based on functional approaches and does not require large basis sets.

\section{Computational Details} 
All CASSCF calculations have been performed using the GAMESS-US program \cite{GAMESS}. The DMFT and $\Pi$DFT calculations have been performed by using a homebrew program that accepts integrals and other quantities from GAMESS-US. The cc-pVTZ (no f functions) basis has been used for all calculations, since this allows us to easily compare the results with a recently published $\Pi$DFT study \cite{PernalGritsenkovanMeer2019}. This choice also allows us to use the parametrization that was used in this study. So
$P^{SDC}(X)$ which governs the suppression of dynamical correlation is given by
\begin{equation}
P^{SDC}(X)= \frac{ax}{1+(a-1)x}
\end{equation} 
with $a = 0.2$, and $P^{EDC}(X)$ which governs the enhancement of dynamical correlation is given by
\begin{equation}
P^{EDC}(X)= c \sqrt[4]{x} -\frac{(c-1)(x-g)^2}{(1-g)^2}
\end{equation} 
with $c = 2.6$ and $g = 1.5$.  

The original parameters of the ELS-DM functional are given by \cite{vanMeerGritsenkoBaerends2018}
\begin{align*}
\alpha_o &= 1.25 \\
\beta_o  &= 750  \\
\gamma_o &= 1500
\end{align*}
These parameters were optimized for reproducing CASSCF energies for CAS spaces of 1 orbital per valence electron. The $\Pi$DFT correction scheme has, in principle, only been applied to smaller active spaces of 2 orbitals per broken bond. In order to facilitate the comparison to earlier CAS$\Pi$DFT application we restrict ourselves to the smaller active space. Simultaneously such a choice allows one to reparametrize 
the original parameters. The following modified parameters have been used to obtain better results in the intermediate bond distance regions for molecules with multiple broken bonds in the same region
\begin{align*}
\alpha_m &= 1.1 \\
\beta_m  &= 250 \\
\gamma_m &= 1500
\end{align*}
In all cases we show the results of both the original (o) and the modified (m) ELS-DM parametrization. 

\section{Results} 
In this section we will show the result of combining the ELS-DM DMFT functional with the $\Pi$DFT scheme for the  H$_2$O (double bond break), N$_2$ and H$_2$CO (C=O bond break) molecules. All of these molecules contain multiple broken bonds. The reason for choosing such a test set is that the exact DMFT functional for 2 electron systems is exactly equal to the CASSCF treatment, and that the ELS-DM functional reduces to the exact functional if one only uses two active electrons, making the comparison between ELS-DM and CASSCF trivial if only single bond breaks were to be discussed. 
 
It should be mentioned that the $\Pi$DFT scheme still has some caveats and it does not always recover all dynamical correlation. The focus of the DMFT-CASSCF comparison is quite reasonable since any corrections to the $\Pi$DFT scheme are more likely to be applicable to both the DMFT and CASSCF methods if all of the initial DMFT and CASSCF quantities are comparable. Below we will look at one of these corrections.

The total energy curves (CASSCF/DMFT + $\Pi$DFT) for a minimal active space of 2 orbitals per broken bond are shown in Figures 1-3. The energy decomposition (CASSCF/DMFT, $\Pi$DFT, CASSCF/DMFT + $\Pi$DFT) of the N$_2$, H$_2$O and H$_2$CO molecules for 3 bond distances (equilibrium, roughly 1.5 times equilibrium (half way dissociated) and roughly 3 times equilibrium (dissociated)) is shown in Table \ref{tab:smallCAS}. 

\begin{figure}[!p] 
\includegraphics[width= 0.7\textwidth, trim= 20 20 20 20]{\figdir/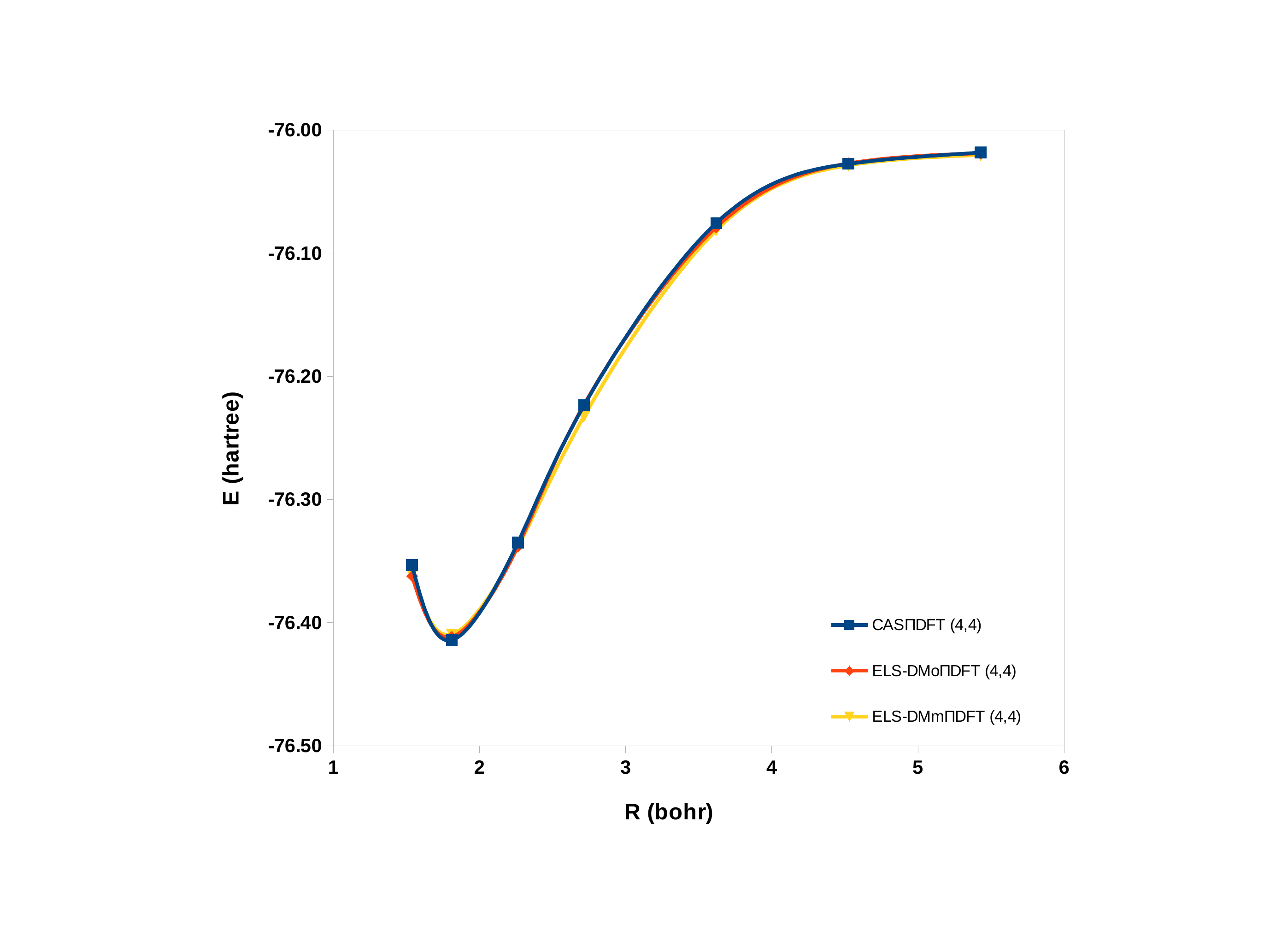} 
\caption{H$_2$O double bond dissociation curves for an active space of 4 electrons in 4 orbitals.} 
\label{fig:H2Oelscas}
\end{figure}

We will begin our analysis with the H$_2$O molecule, for which two "linked but isolated" bonds are dissociated simultaneously. Both of the ELS-DM curves shown in Figure \ref{fig:H2Oelscas} nearly coincide with the CAS$\Pi$DFT curve and Table \ref{tab:smallCAS} shows that the individual components (CAS space and $\Pi$DFT correction) also nearly coincide, indicating that the combined ELS-DM$\Pi$DFT method can act as a substitute for CAS$\Pi$DFT for this molecule. One should note that the modified ELS-DM parametrization yields slightly inferior results compared to the original parametrization. This is not very alarming since the modified parametrization is mainly applicable to situations in which multiple bonds in the same region are broken.

The results for the C=O bond break of the  H$_2$CO molecule paint a similar picture. The ELS-DM$\Pi$DFT energies and their decomposition are close to the CAS$\Pi$DFT ones. However, in this case the variation is slightly higher than for the H$_2$O case, especially at the equilibrium and intermediate bond distances.

The triple bond dissociation of the N$_2$ molecule proves to be a bit more difficult to describe for the ELS-DM functional. The original parametrization of the ELS-DM functional fails in the intermediate bond distance region. We analyzed the CASSCF 2RDM of this region and compared it to the original ELS-DM results and noted that the dispersive type interactions were present in CASSCF, while they were nearly absent for ELS-DM. The modified parameter set fixes this issue and yields good overall results.

We have seen that the ELS-DM$\Pi$DFT method is, after some reparametrization, capable of reproducing the CAS$\Pi$DFT results. However, as was mentioned before, the CAS$\Pi$DFT method is not without its errors. It is has been shown that the dissociation limit of CAS$\Pi$DFT for multibonded systems is too high compared to the accurate CBS limit \cite{PernalGritsenkovanMeer2019,HapkaPernalGritsenko2020}. The main culprit is the lack of the interbond dynamical correlation between the electrons localized on the same fragment of a dissociating molecule. Two different correction schemes were proposed. The newest variant injects the density of the frontier orbitals into a LYP functional scaled by an occupation number dependent prefactor. The older variant, which we will be using here, essentially entails using the correlation correction for multibond systems eq \ref{eq:multibond} again. The resulting CAS$\Pi$DFT+M method
\begin{align}
&E_e^{CAS\Pi DFT+M} =  E_e^{CAS\Pi DFT} \nonumber \\&+  \int d\boldsymbol{r}_1 d\boldsymbol{r}_2 \frac{\Pi^{\text{M}}[\gamma^{\text{CASSCF}}(\boldsymbol{r},\boldsymbol{r}')](\boldsymbol{r}_1,\boldsymbol{r}_2)}{|\boldsymbol{r}_1-\boldsymbol{r}_2|}
\end{align}
has shown to be capable of relatively accurately reproducing the CBS curve for the N$_2$ molecule. Similarly, one can add the correction (again) to the ELS-DM$\Pi$DFT method
\begin{align}
&E_e^{ELS-DM\Pi DFT+M} =  E_e^{ELS-DM\Pi DFT} \nonumber \\&+ \int d\boldsymbol{r}_1 d\boldsymbol{r}_2 \frac{\Pi^{\text{M}}[\gamma^{\text{ELS-DM}}(\boldsymbol{r},\boldsymbol{r}')](\boldsymbol{r}_1,\boldsymbol{r}_2)}{|\boldsymbol{r}_1-\boldsymbol{r}_2|}
\end{align}
resulting in a method that is ought to be able to reproduce the N$_2$ curve. As is shown in Figure \ref{fig:N2elscasM}, this is indeed the case, proving that the ELS-DM functional is a satisfactionary replacement for the CASSCF wavefunction for ground state calculations.

\section{Conclusions} 
A combined approximate density and density-matrix ELS-DM$\Pi$DFT(+M) functional method is proposed for calculation of potential energy curves (PECs) of molecular multibond dissociation. It accounts for all relevant effects of electron correlation along the bond dissociation coordinate. These effects include the short-range dynamical correlation, the long-range intrabond non-dynamical correlation, as well as the important in the dissociation region medium range interbond electron correlation.

The key point of the present DMFT+DFT development is that the ELS-DM pair density, a relatively simple 1RDM functional, closely reproduces a more complicated {\it ab initio} pair density of CASSCF. This allows the corresponding ELS-DM energy functional to efficiently account for non-dynamical correlation.

Furthermore, the ELS-DM on top pair density closely reproduces locally the CASSCF on top pair density. This allows to physically meaningfully correct the correlation DFT LYP functional for SDC using the generated within DMFT on top pair density within $\Pi$DFT.

The proposed ELS-DM$\Pi$DFT functional is applied to calculation of the PECs of the multibond dissociation in the prototype molecules N$_2$, H$_2$O, and H$_2$CO. The resultant ELS-DM$\Pi$DFT PECs go very close to the corresponding PECs of the CAS$\Pi$DFT method, which has been recently successfully applied to the calculation of various molecular PECs in Refs. \cite{GritsenkovanMeerPernal2018,GritsenkovanMeerPernal2019,PernalGritsenkovanMeer2019,HapkaPernalGritsenko2020}.

The proposed ELS-DM$\Pi$DFT effectively resolves the major DMFT bottleneck, stemming from the troublesome feature of the 1RDM spectrum, namely, the accumulation of the NO eigenfunctions near the zero NON eigenvalue. Because of this feature, reaching the fully self-consistent solution with a DM functional, which includes all NOs in a given basis, often becomes, in a general case, a veritable numerical nightmare. 

The present ELS-DM$\Pi$DFT functional efficiently circumvents this DMFT bottleneck by not using at all the higher NOs outside the minimal geminal subsets. In conventional DMFT the inclusion of these NOs is required to properly account for dynamical correlation. At variance with this, in ELS-DM$\Pi$DFT dynamical correlation is evaluated with the $\Pi$DFT functional, which does not use higher NOs. With the results obtained, this can be considered as a further development in the functional theory focused on the reliable calculation of the molecular PECs.

\begin{acknowledgments} 
This work was supported by the Ministry of Science and Technology of Taiwan (Grant No.\ MOST107-2628-M-002-005-MY3), National Taiwan University (Grant No.\ NTU-CDP-105R7818), and the 
National Center for Theoretical Sciences of Taiwan. 

O. G. gratefully acknowledges the support by the Narodowe Centrum Nauki of
Poland under Grant No. 2017/27/B/ST4/00756.
\end{acknowledgments} 

\bibliography{dmftpidftBIB.bib}

\newpage

\begin{figure}[!p] 
\includegraphics[width= 0.7\textwidth, trim= 20 20 20 20]{\figdir/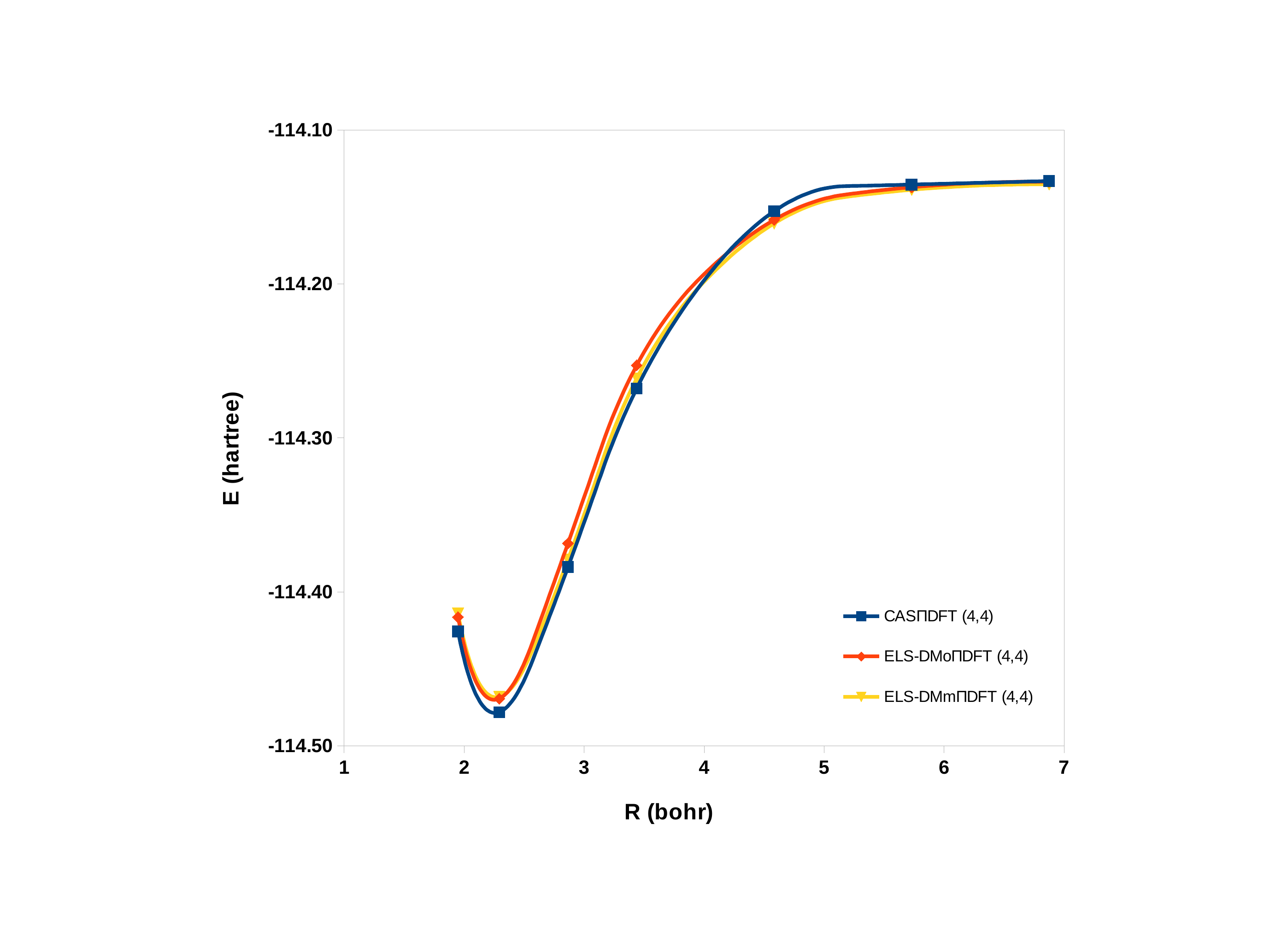} 
\caption{H$_2$CO C=O double bond dissociation curves for an active space of 4 electrons in 4 orbitals.} 
\label{fig:H2COelscas}
\end{figure} 

\begin{figure}[!p] 
\includegraphics[width= 0.7\textwidth, trim= 20 20 20 20]{\figdir/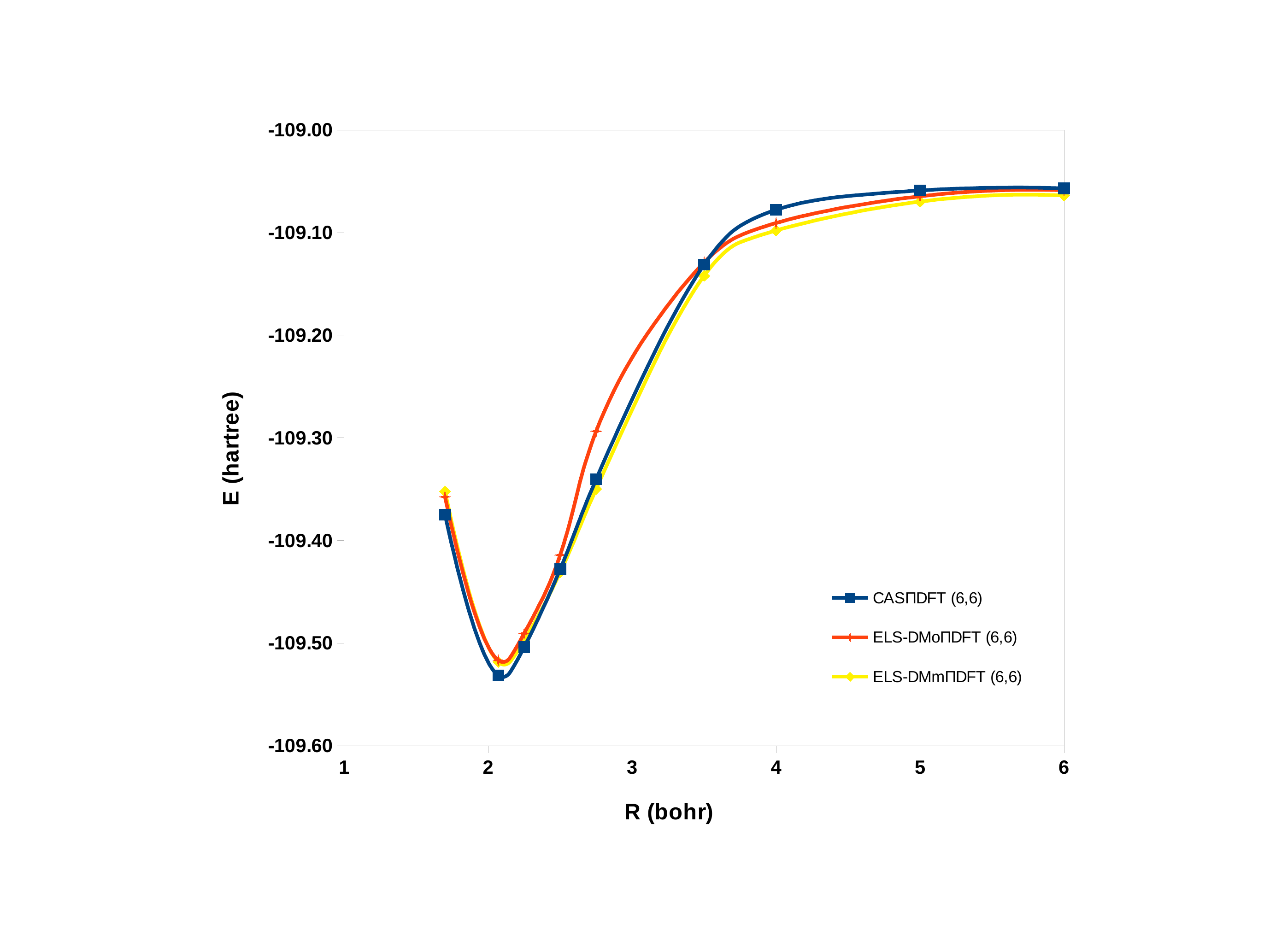} 
\caption{N$_2$ triple bond dissociation curves for an active space of 6 electrons in 6 orbitals.} 
\label{fig:N2elscas}
\end{figure} 

\begin{figure}[!p] 
\includegraphics[width= 0.7\textwidth, trim= 20 20 20 20]{\figdir/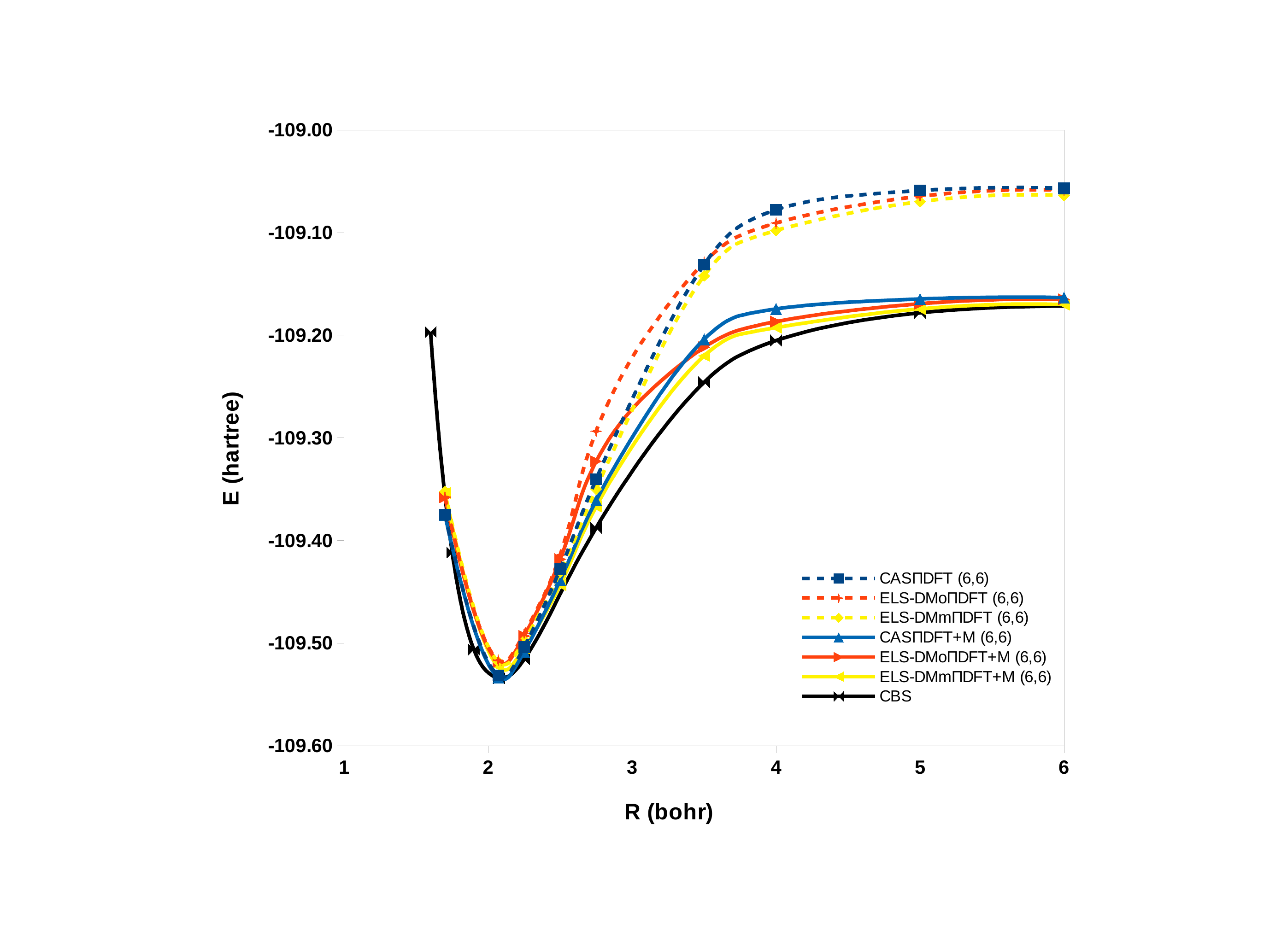} 
\caption{N$_2$ triple bond dissociation curves for an active space of 6 electrons in 6 orbitals. The dashed curves represent the calculations without the medium range correlation dissociation correction. The CBS data has been taken from Ref. \cite{LieClementi1974B}.} 
\label{fig:N2elscasM}
\end{figure} 

\begin{table}[!p]
\centering
\begin{tabular}{l | l | r r r | r r r| r r r} 
	&	&	&N$_2$	&	&	&H$_2$CO	&	&	&H$_2$O	&\\
R (bohr)	&	&2.075	&2.75	&6.0	&2.292	&3.468	&6.876	&1.81	&2.72	&5.43\\
\hline
	&CASSCF	&-109.1166	&-108.9732	&-108.7949	&-113.9816	&-113.8219	&-113.7331	&-76.1084	&-75.9549	&-75.8057\\
CAS space	&ELS-DMo	&-109.1088	&-108.9455	&-108.7966	&-113.9749	&-113.8188	&-113.7336	&-76.1100	&-75.9564	&-75.8064\\
	&ELS-DMm	&-109.1222	&-108.9909	&-108.8019	&-113.9781	&-113.8255	&-113.7354	&-76.1090	&-75.9664	&-75.8079\\
\hline
	&CASSCF	&-0.4148	&-0.3671	&-0.2620	&-0.4966	&-0.4460	&-0.4000	&-0.3058	&-0.2687	&-0.2127\\
$\Pi$DFT	&ELS-DMo	&-0.4078	&-0.3480	&-0.2621	&-0.4944	&-0.4340	&-0.4000	&-0.3014	&-0.2667	&-0.2125\\
	&ELS-DMm	&-0.3958	&-0.3587	&-0.2620	&-0.4901	&-0.4359	&-0.3999	&-0.3005	&-0.2661	&-0.2124\\
\hline
	&CASSCF	&-109.5314	&-109.3402	&-109.0568	&-114.4782	&-114.2679	&-114.1331	&-76.4142	&-76.2236	&-76.0184\\
CAS + $\Pi$DFT	&ELS-DMo	&-109.5166	&-109.2935	&-109.0587	&-114.4693	&-114.2528	&-114.1336	&-76.4115	&-76.2230	&-76.0189\\
	&ELS-DMm	&-109.5181	&-109.3496	&-109.0638	&-114.4682	&-114.2614	&-114.1353	&-76.4095	&-76.2325	&-76.0203\\
\end{tabular}
\caption{Energies in hartree for minimal active space calculations. The CAS section shows the CASSCF energies and the DMFT energies that try to approximate it. The $\Pi$DFT section gives the dynamical correlation correction. The CAS + $\Pi$DFT section shows the sum.}
\label{tab:smallCAS}
\end{table}

\end{document}